\documentclass[pre,superscriptaddress,showpacs,nofootinbib,twocolumn,notitlepage]{revtex4-1} 

\usepackage{amssymb}   
\usepackage{amsmath}   
\usepackage{natbib}
\usepackage{epsfig}   
\usepackage{graphicx,subfigure}   
\usepackage{dcolumn}
\usepackage{bm}  
\usepackage[english]{babel}
\usepackage[latin1]{inputenc}
\usepackage{eucal} 
\usepackage{verbatim}
\usepackage{latexsym}

\usepackage{xcolor}

\usepackage[update]{epstopdf}
\begin{document}

\title{Environment  fluctuations on single species pattern formation}
\author{L. A. da Silva} \email{leandro.silva@ufabc.edu.br}
\affiliation{Center of Mathematics, Computation and Cognition, UFABC, S\~ao Paulo, Brazil}
  
\author{E. H.  Colombo} \email{eduardo.colombo@fis.puc-rio.br}
\affiliation{Department of Physics, PUC-Rio, Rio de Janeiro, Brazil}

\author{C. Anteneodo} \email{celia.fis@puc-rio.br}
\affiliation{Department of Physics, PUC-Rio, Rio de Janeiro, Brazil}
\affiliation{Institute of Science and Technology for Complex Systems, Rio de Janeiro, Brazil}

\begin{abstract}

System-environment interactions are intrinsically  
nonlinear and dependent on the interplay between many degrees of freedom. 
The complexity may be even more pronounced when one aims to describe biologically motivated systems.  
In that case, it is useful to resort to simplified models relying on effective stochastic equations. 
A natural consideration is to assume that there is a  noisy contribution from the 
environment, such  that the parameters  which characterize it are not constant but instead fluctuate 
around their characteristic  values. 
From this perspective, we propose a stochastic generalization of the nonlocal Fisher-KPP equation where, 
as a first step, environmental fluctuations are Gaussian white noises, both in space and time. 
We apply analytical and numerical techniques to study how noise affects stability and pattern formation in this 
context. 
Particularly, we investigate  noise induced coherence by means of the complementary information 
provided by the dispersion relation and the structure function.

\end{abstract}


\maketitle

\section{Introduction}

The mathematical description of the spatial distribution of biological populations 
can be achieved on a phenomenological mesoscopic level 
where system and environment properties are typified by means of a few control parameters. 
The evolution of a population distribution 
is mainly ruled by processes such as reproduction \cite{Malthus1826} and 
(interspecific or intraspecific)  competitions, which are  usually mimicked by  logistic-like expressions \cite{Verhulst1838}, 
together with  spatial dispersal,   modeled by (normal or anomalous) diffusion. 
Then, the population characteristics and the coupling to the environment are quantified by a set of control parameters, 
such as growth rate, carrying capacity and diffusion coefficient, each one assigned a typical value. 
Such simple models allow to predict the relaxation towards a steady state, resulting from the interplay between 
the population growth and the competition for resources in the limited support provided by the environment. 
However, the long-time evolution of biological populations can present complex spatiotemporal patterns,  a signature of self-organization, 
as can be observed in populations of slime mold,  bacteria, ants, birds,  
fishes and  human beings \cite{ants,birds,bacteria,mold,human}.
Self-organization  may arise due to nonlocal 
interactions~\cite{murray2002mathematical,Fuentes2004,Fuentes2003,DaCunha2011,Allee} 
or other mechanisms that drive the system far from equilibrium towards a spatiotemporal organization. 
As an example,  in a recent study about the \textit{Allee effect}~\cite{reviewAllee}, 
it has been shown that the interplay between  nonlocality and 
 nonlinearity can lead to the emergence of localized structures~\cite{Allee}.

The environment certainly interferes in most of those processes. 
For example,  for microorganisms,  the environment temperature  can affect the 
reproduction rate \cite{Ratkowsky1982} and many other processes \cite{micro} such as spatial spread. 
Competition  is  intrinsically mediated by the environment due to 
its limited  resource availability (carrying capacity) \cite{Verhulst1838}. 
Now, due to the inherent complexity,  an environment parameter is typically subjected to a complex 
web of diverse processes, 
 varying at different scales, both in space and time.  
Therefore,  it would be more realistic to model its complicated behavior by means of  a stochastic variable.  
It is our goal to investigate the impact of such fluctuations on  population dynamics.
We will consider a single species scenario in one-dimension for the evolution of the population density $u(x,t)$. 

A standard deterministic model that takes into account the above mentioned governing rules  
is the  generalized Fisher-KPP equation \cite{Kolmogorov1937,fisher1937wave,Fuentes2003}, 
namely, the adimensionalized integro-differential equation 
\begin{equation}\label{maineq}
\frac{\partial u(x,t) } {\partial t}   = a\,u(x,t)  - b\,u(x,t)\,\mathcal{J}[u](x,t) + D\frac{\partial^2 }{\partial x^2} u(x,t)  \,,
\end{equation}
where  $\{a,b,D\}$ are positive parameters and 
$\mathcal{J}[u](x,t) = \int_\Omega f( x- x') u( x',t)d x'$, with $f$ a function that 
describes the influence of two interacting infinitesimal elements at a distance $x-x'$. 
The first term in Eq.~(\ref{maineq}) accounts for the balance between the birth and death rates and the 
second one introduces a nonlocal intraspecific competition that sets a saturation 
limit on  population   growth. Then they can be seen as a generalization of the Verhulst expression.  
The last term introduces spatial spread through normal diffusion~\cite{murray2003mathematical}. 

In Eq.~(\ref{maineq}), the environment participates in defining all the 
set of control parameters $\{a,b,D\}$.
The inclusion of small fluctuations (or noise)   allows to reflect  
the spatiotemporal variability of the complex environment. 
We   focus on the effects of  the multiplicative noise 
that arises by resorting to the transformation $a  \rightarrow a + \sigma_{\eta} \eta( x,t)$, 
and we also consider the impact of an additive noise $\sigma_{\xi} \xi(x,t)$, 
where $\sigma_{\eta}$ and $\sigma_{\xi}$ 
are constant parameters that control the amplitude of the fluctuations, 
and where   both $\xi$ and $\eta$ are independent Gaussian noises,  with null averages 
$\langle \eta( x,t)  \rangle = \langle \xi( x,t)  \rangle = 0$,  
and white in space-time, i.e., 
$\langle \eta( x,t)\eta( x',t')  \rangle = \delta(x-x')\delta(t-t')$ 
and $\langle \xi( x,t)\xi( x',t')  \rangle = \delta(x-x') \delta(t-t')$. 
In the context of population dynamics these multiplicative and additive 
fluctuations introduce the complex aspects of the environment 
in the growth rate and in the flux of individuals  through the system boundaries, respectively. 

Therefore, our object of study is the dynamical equation that can be cast in the following form:
\begin{eqnarray} 
\frac{\partial u(x,t)}{\partial t}  &=&  {\bigl( a+ \sigma_{\eta}\eta( x,t) \bigr)  }  
u( x,t) +  \sigma_{\xi} \xi(x,t) +  \nonumber\\
&&   - b  u(x,t) \mathcal{J}[u]  + D \frac{\partial^2 }{\partial x^2} u( x,t)\;. \label{nonlocalfisherlangevin}
\end{eqnarray}
Since the shape of the influence function does not lead to substantially 
different results \cite{Fuentes2003}, for the sake of simplicity, 
 throughout this work  we will use a Heaviside influence function defined as 
$f( x-x') = \frac{1}{2 w} \Theta(w - | x- x'|)$. 
In the previous expression, $w$ is a positive constant, defining the range of 
the interactions. 
 
Moreover, for the multiplicative white noise term, one must state an additional prescription 
(typically, either It\^o or Stratonovich) \cite{vankampen}. 
Within the present scenario, there are cases in which the It\^o interpretation is suitable,  
for example 
i) when the environment is sensed in a nonantecipative manner by the individuals \cite{Turelli},  
ii)  when the continuous model is actually an approximation for a discrete time population 
evolution \cite{lande}, or iii) when fluctuations originate from internal sources. 
On the other hand, if fluctuations are external,  the Stratonovich description is more appropriate. 
This is so because in the latter case   the deterministic drift is recovered in the limit of 
vanishing fluctuations, while the drift is coupled to the fluctuations when 
they have internal origin \cite{vankampen}.  
Since both rules may be sound depending on the specific environment fluctuations 
taken into account, we will consider both of them, making a comparison of their different 
effects while the model parameters are kept the same. 
This is the way we chose to present the results,
despite formally a correspondence between both descriptions will ultimately amount 
to the modification of the deterministic term.

The effects of additive noise in spatially extended systems
have been tackled by previous works (for example see Refs.~\cite{Sagues2007,additive,book}). 
Now, we analyze these effects in the context of population dynamics.
The additive noise accounts for the effect of a system-environment
coupling that is density-independent. 
Additive fluctuations represent fluctuating fluxes of individuals through the system's boundary.
Then, even when the population tends to vanish, a positive
fluctuation represents a reintroduction of individuals
into the system, spoiling the absorbent state. 
A negative value of the noise reflects a tendency to an outwards flux. 
But, this can be accomplished only until a null value of $u$ is attained, 
then in numerical solutions, at each step,  we trimmed fluctuations that would lead to a negative value of $u$.
In fact,  additive noise in Eq.~(\ref{nonlocalfisherlangevin})
compromises the positivity of the population density, therefore  
the stochastic equation must be complemented 
by an additional mathematical constraint, as the one chosen, to forbid negative values of $u$.

\section{Instability conditions}

\begin{figure}[h!]
\includegraphics[scale=0.6]{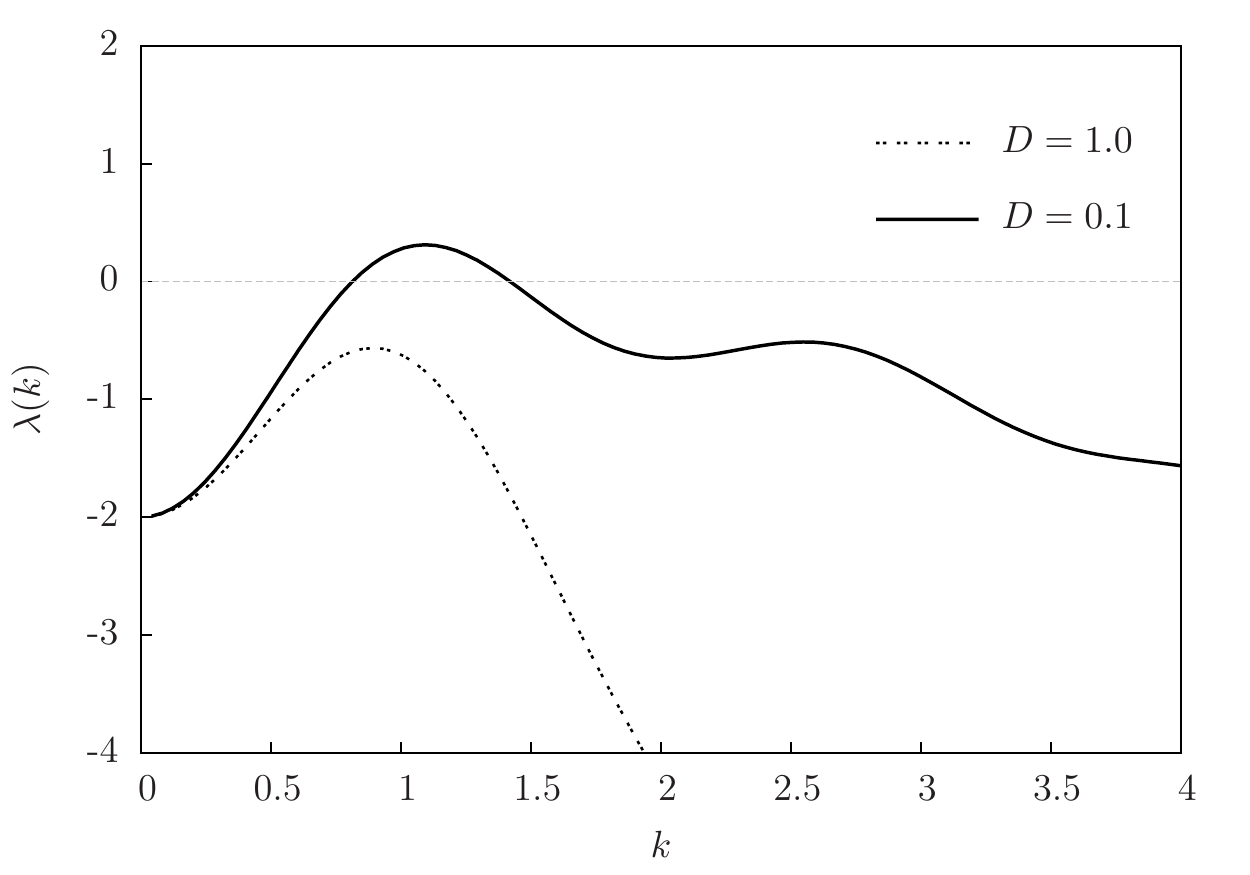}
\caption{Dispersion relation given by Eq.~(\ref{detDispersion}), for $a=2$, $b=1$, $w=4$, and 
two values of the diffusion coefficient $D$ indicated on the figure.}
\label{fig:dispersion}
\end{figure}

In the  deterministic case \cite{Fuentes2003}, i.e. when $\sigma_\xi=\sigma_\eta = 0$,  
one can determine the instability condition for the emergence of periodic structures 
by following the standard procedure of linearizing Eq.~(\ref{maineq}) 
around the homogeneous solution $u_0 =a/b$, 
assuming  $u(x,t) = u_0 + \varepsilon(x,t)$, where $\varepsilon(x,t)=\varepsilon_0 \exp[i k x+\lambda(k)t]$ 
is a small perturbation around the uniform state  $u_0$. 
This procedure leads to the dispersion relation
\begin{equation}\label{detDispersion}
\lambda(k) =  - a\tilde{f}(k) -Dk^2\,,
\end{equation}
where $\tilde{f}$ is the Fourier transform of the influence function, that in the 
particular case of the Heaviside influence becomes $\tilde{f}(k)= \sin{(w k)}/[w k]$. 
The relation (\ref{detDispersion}) indicates  instability 
with respect to a certain mode $k$, if $\lambda(k)>0$. 
Then, in general, patterns are expected if the dispersion relation satisfies two conditions: 
(i) $\lambda(0) < 0$, to avoid instability of the average population size, 
and (ii) there must exist a positive global maximum at certain $k^* >0$ \cite{Cross1993}, 
to give rise to an emergent characteristic mode. 
Recently, it has been shown that this relation provides important information about the pattern formation process 
not only for  short times but also asymptotically, as soon as mode coupling is weak~\cite{Colombo2012}. 
In particular, diffusion has a stabilizing role, while the first term has not a definite sign. 
Following the dispersion relation (\ref{detDispersion}),   
we show  in Fig.~\ref{fig:dispersion} that,
when the diffusion coefficient is reduced with the other parameters kept constant,   
the homogeneous solution can become unstable  and patterns  emerge 
in the population \cite{Fuentes2003,Colombo2012}. 
Both instances are depicted in Fig.~\ref{fig:dispersion},   
for fixed parameters $a, b, w$. Notice that in both cases $\lambda(0)<0$, but, for small $D$, 
$\lambda(k)$ takes positive values, while for $D$ above a threshold value 
$\lambda(k)$ is always negative indicating the stability 
of the homogeneous state. 
 
Now let us turn to the stochastic version of the nonlocal Fisher-KPP equation. 
By linearizing Eq.~(\ref{nonlocalfisherlangevin}) around $u_0=a/b$, 
in the small noise approximation, we have
\begin{align} \label{perturbation}
\frac{\partial \varepsilon(x,t)}{\partial t} &= \;  -a\mathcal{J}[\varepsilon](x,t)   
+ D \frac{\partial^2}{\partial x^2} \varepsilon(x,t)  +\nonumber \\
& + \sigma_{\eta} \varepsilon(x,t)  \eta(x,t)   + \sigma_{\eta} u_0\eta(x,t) +\sigma_{\xi} \xi(x,t)\,,
\end{align}
where the deterministic terms  are represented in the first line of the right hand side 
of Eq.~(\ref{perturbation}), while  the second line contains 
the multiplicative and additive noise terms.
A suitable way to verify pattern formation is to  measure the spatial autocovariance 
$C(r,t)=\int \langle \varepsilon(x,t)\varepsilon(x+r,t) \rangle dx$ (which does not depend on $t$ if stationarity holds) or, 
alternatively, its Fourier transform, that is the structure function
\begin{equation}
 S(k,t) \equiv \left\langle \hat{\varepsilon}(k,t) \hat{\varepsilon}(- k,t) \right\rangle \,,
\end{equation}
where $\hat{\varepsilon}$ is the Fourier transform of $\varepsilon$.

Following the lines of Refs. \cite{book,stability}, 
we derive the evolution equation of $S(k,t)$ under the Stratonovich interpretation. 
Starting from the Fourier transform of Eq.~(\ref{perturbation}), 
 considering $\partial_t(\hat{\varepsilon}\hat{\varepsilon}^\prime)=
\hat{\varepsilon}\partial_t(\hat{\varepsilon}^\prime)+\hat{\varepsilon}^\prime\partial_t(\hat{\varepsilon})$, 
where $\hat{\varepsilon}  \equiv \hat{\varepsilon}( k,t)$ and 
$\hat{\varepsilon}^\prime \equiv \hat{\varepsilon}(-k,t)$, and averaging, 
 we obtain
\begin{equation}   \label{Sprelim}
\frac{1}{2}\frac{\partial}{\partial t} S(k,t)   = \lambda(k) S(k,t) 
 +     \sigma_{\eta} \langle \hat{\varepsilon}^\prime \widehat{\varepsilon\eta} \rangle
+     \sigma_{\eta} u_0 \langle \hat{\varepsilon}^\prime \hat{\eta} \rangle
 + \sigma_{\xi}  \langle \hat{\varepsilon}^\prime  \hat{\xi}\rangle
\,.
\end{equation}  

In order to evaluate the average of multiplicative terms, we resort, 
for the case of the Stratonovich interpretation, 
to the so-called Furutsu-Novikov theorem \cite{novikov}, namely 
\begin{equation} \label{furutsu-novikov0}
\langle \chi(q) g[\chi] \rangle =  \int dy \langle \chi(q) \chi(y) \rangle 
\left\langle \frac{\delta\left[g(q)\right]}{\delta \chi(y)}   \right\rangle\,,
\end{equation}  
where $g(q')$ is functionally dependent on the Gaussian stochastic process $\chi$. 
Hence, the averages of interest,  in the small noise approximation,  are 
\begin{align}
2\langle \hat\varepsilon^\prime\widehat{\varepsilon \eta} \rangle &= 
 \sigma_\eta     K_\eta S(k,t) \, , \\
2\langle \hat\varepsilon^\prime  \hat\eta   \rangle &=  \sigma_\eta u_0 \, , \\
2\langle \hat\varepsilon^\prime  \hat\xi    \rangle &=  \sigma_\xi  \, ,
\end{align}
where  $K_\eta$ is to be interpreted as the spatial correlation function 
of the noise  for $x=x'$,  numerically computed as 
$ K_{\eta}=1/\Delta x$, where $\Delta x$ is the lattice spacing. 

Finally, substituting the  averages into Eq.~(\ref{Sprelim}),
the  dynamical equation for the structure function reads 
%
%
\begin{align} \label{structure_dynamics}
\frac{\partial}{\partial t} S(k,t)   = 2\Lambda_\nu(k) S(k,t) 
 +     \sigma_{\eta}^2  u_0^2 + \sigma_{\xi}^2   \,,
\end{align}
with 
\begin{equation} \label{stocDispersion0}
\Lambda_\nu(k) =   - a \tilde{f}(k) - D k^2 + 
\frac{1}{2}  \nu   \sigma_{\eta}^2K_{\eta} \,,  
\end{equation}
where, the factor $\nu$  allows to select either the It\^o ($\nu=0$) or Stratonovich ($\nu=1$) rules. 
The former case is obtained either by including the spurious drift to the original equation to transform 
the noise into a Stratonovich one or simply by considering that the average of multiplicative noise terms 
vanish. 

Notice that Eq.~(\ref{stocDispersion0}) can be identified as the stochastic generalization
 of the dispersion relation given by Eq.~(\ref{detDispersion}).
 If $\Lambda_\nu(k)$ is positive in some range of $k$, 
then  perturbations grow,  indicating that the homogeneous state $u_0$ is unstable. 
Otherwise, i.e., if $\Lambda_\nu(k)<0$ for all $k$, the state $u_0$ is stable and perturbations vanish. 
The contribution of noise is given by the last term in  Eq.~(\ref{stocDispersion0}), that is 
always nonnegative and independent on $k$. No such effect is predicted when noise is interpreted 
under the It\^o rule ($\nu=0$). 
In any case, the additive noise does not affect the dispersion relation. 
Also notice that, although the multiplicative noise $\eta$  is destabilizing  in the Stratonovich case,   
it will affect all modes. 
Then, the dispersion relation obtained by the linear analysis already points out that noise can reveal the  
  instability built by the nonlocal competitive interactions. 

Additional information can be obtained from the structure function. 
Under stationarity,  Eq.~(\ref{structure_dynamics}) leads to    
\begin{equation} \label{structure}
  S(k)   =  \frac{\sigma_{\eta}^2  u_0^2 + \sigma_{\xi}^2}{	- 2\Lambda_\nu(k) } \,.
\end{equation}
  
Therefore, although the analysis of the signal of $\Lambda_\nu(k)$ predicts 
no effects caused by noise under the It\^o prescription, 
the structure function reveals that noise can induce some kind of coherence. 
The numerical analysis in the next sections will clarify this issue.

However, let us remark that  the stationary amplitude, $S^\star\equiv S(k^\star)$, 
of the dominant mode $k^\star$  grows with both noise intensities. 
Moreover, note that $k^\star$ is defined by the deterministic component only, 
hence by the  dispersion relation (\ref{detDispersion}). It is in that sense   
that noise reveals the instability of a hidden dominant mode that has been built  
by the nonlocal interactions and suppressed by the homogenizing diffusion process. 
It is also noteworthy that, according to Eq.~(\ref{structure}), noise 
has a constructive role only if $u_0 > 0$.  
 The role of noise as a precursor phenomenon and 
as a factor that induces coherence  is already 
known in other dynamical systems~\cite{Sagues2007,schopf1993convection,sancho,additive}. 
In the following section we analyze the impact of  noise in the context 
of Eq.~(\ref{nonlocalfisherlangevin}).

\section{Impact of noise}

The above analytical statements  allow to predict the stability of  
the homogeneous state in the presence of noise in  the dynamic rules. 
That analysis  tacitly assumes the stability of the homogeneous distribution in the absence 
of noise, i.e., $\lambda(k )<0$ for all $k$, a situation that  will be numerically 
investigated in subsection  \ref{sec:h}. 
We study the impact of  noise on  deterministically induced patterns ($\lambda(k^\star)>0$), 
 in subsection  \ref{sec:nonh}. 

In order  to go beyond the small noise and linear approximations 
and shed light on the far from equilibrium and nonlinear dynamics, 
 we perform  numerical integration of Eq.~(\ref{nonlocalfisherlangevin}).
We follow the Heun algorithm for stochastic equations \cite{Sagues2007}, 
discretizing space and time, with $\Delta x=10^{-1}$ and $\Delta t < 10^{-3}$. 
We use a one-dimensional array, to represent a system of size $L=100$, with periodic boundary conditions.

\begin{figure}[b!]
\includegraphics[scale=0.6]{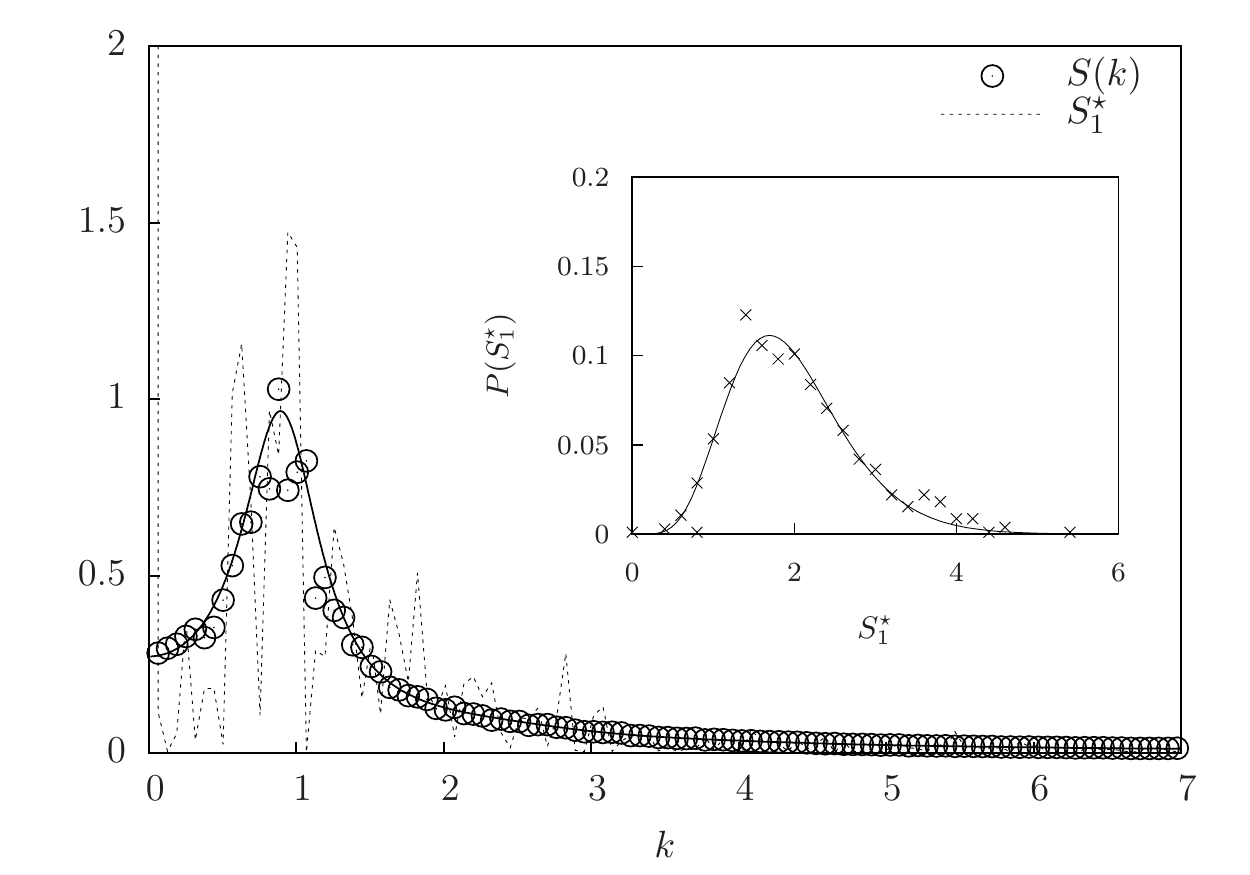}
\caption{Stationary structure function $S(k)$,  for $a=2$, $b=1$, $w=4$, $D=1$, 
$\sigma_\xi=0$ and $\sigma_{\eta}=0.5$ (It\^o noise), 
obtained numerically (symbols) and through Eq.~(\ref{structure}) with $\nu=0$ (solid line). 
For comparison, we also display the stationary power spectrum for an individual realization $S_1(k)$ (dotted line). 
The inset shows the distribution of values of $S_1^\star=S_1(k^\star)$: numerical (symbols) and 
gamma distribution fit  as a guide to the eyes (solid line). 
}
\label{fig:structure}
\end{figure}

In all cases, we quantify spatial coherence, at a given time $t$, 
by means of the structure function, which is an ensemble average. 
Averages over 100 samples were considered.
From the structure function, one can extract the dominant mode  $k^\star$ and its corresponding amplitude.  
After a transient period, the stationarity of the structure function is attained. 
The stationary characteristic mode  is well predicted by Eq.~(\ref{structure}), as 
illustrated in Fig.~\ref{fig:structure} where we show a comparison between the numerical result 
for the stationary structure function and the linear theory prediction for the It\^o case, given by 
Eq.~(\ref{structure}) with $\nu=0$.  
For Stratonovich, the scenario is qualitatively similar,  as soon as the noise intensity is 
small enough. 
Through numerical simulations, one can observe  that the dominant mode $k^\star$  
adopts a typical value, in the whole noise intensity range, 
showing that the uncorrelated noise introduced in the dynamics appears  in a correlated manner. 

The  maximum value $S^\star=S(k^\star)$ gives a measure of the intensity of the dominant mode. 
In the inset of Fig.~\ref{fig:structure} we show a normalized histogram of $S^\star_1$ for 
an individual realization. 
Then, although there exists a good agreement between the numerical structure function 
and its theoretical prediction, there is a large dispersion as depicted by means of an individual realization 
(dotted line) and also by the distribution of values of $S_1^\star=S_1(k^\star)$, 
where $S_1(k)$  is the power spectrum of each single realization.

 The structure function $S$ provides full information about spatial coherence. 
We use its maximum value $S^\star$  to quantify coherence through a single variable, 
but  some warnings are required for a proper interpretation. 
We remark that, even when  $S^\star$  allows to detect a characteristic scale, 
at high noise intensity,  $S$ tends to flatten as more modes become activated, then  patterns become noisier. 
Furthermore,  the structure function measures coherence in Fourier space, 
and it does not guarantee that  patterns are persistent in position space. 
In subsection \ref{sec:h} we shall explore this aspect, by directly measuring the temporal 
correlation and its dependency on noise intensity.

\subsection{On deterministically induced patterns ($\lambda(k^\star)>0$) }
\label{sec:nonh}

Let us consider values of the parameters for which patterns arise in the absence of noise. 
Then, we use for instance the same values of  the  solid line in Fig.~\ref{fig:dispersion}.
We discuss spatial coherence  as a function of noise intensity mainly through the steady value  of $S^\star$. 
We also observe  the spatial average density $\langle u \rangle$.
By means of numerical simulations, we will go beyond the linear approximation to explore the large fluctuations limit.
We first analyze the effect of multiplicative noise (with $\sigma_\xi=0$),  
under both It\^o and Stratonovich prescriptions.

\begin{figure}[b!] 
\includegraphics[scale=0.6]{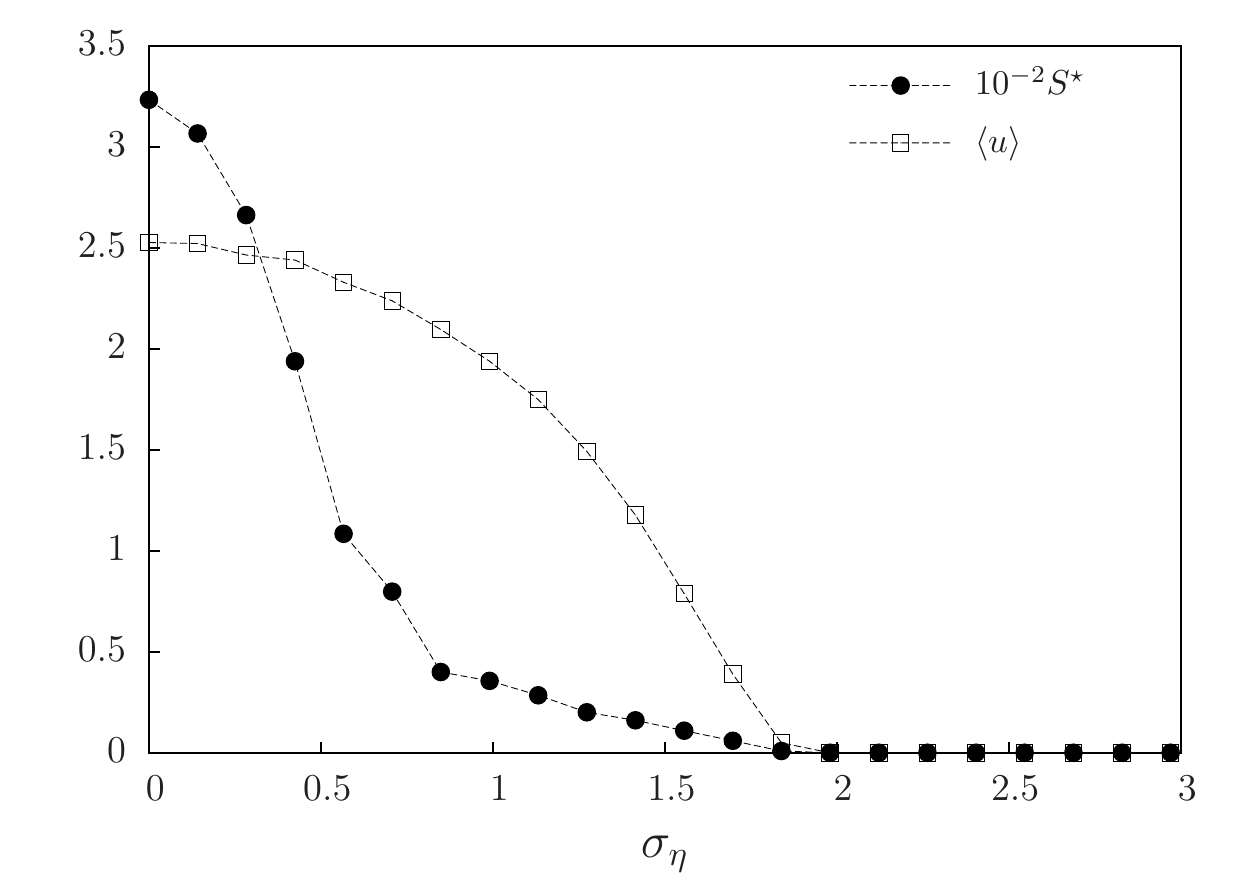}
\caption{Analysis of the spatial coherence. Steady values of the intensity of the 
dominant mode  $S^\star =S(k^\star)$ 
and of the average population density, $\langle u \rangle$, as a function of $\sigma_\eta$ (under the It\^o  prescription), 
for the choice: $a=2$, $b=1$, $w=4$, $D=0.1$ and $\sigma_\xi=0$. 
The intensity of the dominant mode was scaled 
by a factor $10^{-2}$ just to employ a unique axis scale. The dotted lines are  guides to the eyes. 
}
\label{fig:noise_com_I}
\end{figure}

Let us start by the It\^o case. In Fig.~\ref{fig:noise_com_I},  
we represent  the steady value of the spatial average density $\langle u \rangle$, 
together with $S^\star$. 
 The results  show that multiplicative noise plays a destructive role.
Both  quantities decay with noise intensity $\sigma_\eta$. 
Moreover, for the set of parameters   chosen, there exists a threshold value, 
$\sigma_{\eta}^e \approx 1.8$ in the case of the figure, that represents 
the \textit{extinction threshold}. This means that, for noise intensities 
greater than the threshold, the population becomes extinguished. 
This implies a shift transition \cite{horsthemke2006noise} for the critical growth rate 
that now competes with noise. 
This effect may be attributed to the fact that 
the effective growth rate $a + \sigma_{\eta} \eta$ can take negative values. 
In such case, the shape of the effective drift potential changes so that the null 
state can become  instantaneously stable. 
This becomes more frequent as the noise amplitude becomes comparable to  $a$,  
then  creating  a bias towards low densities until extinction.
It is noteworthy that, in the  local mean-field approximation described by the equation
$ du/dt = (a-bu)u + \sigma \eta u$,  
the ensemble average stationary density is given by $\bar{u} = a/b -\sigma^2/(2b)$, 
indicating the existence of a  critical threshold. 
When $\lambda(k^\star)>0$, 
through the deterministic mechanisms the system would  go towards a stationary state that is represented 
by a well defined population distribution pattern, meanwhile the presence of multiplicative 
noise in the dynamic forces, even at low intensity, spoils that spatial order.

\begin{figure}[h!] 
\includegraphics[scale=0.6]{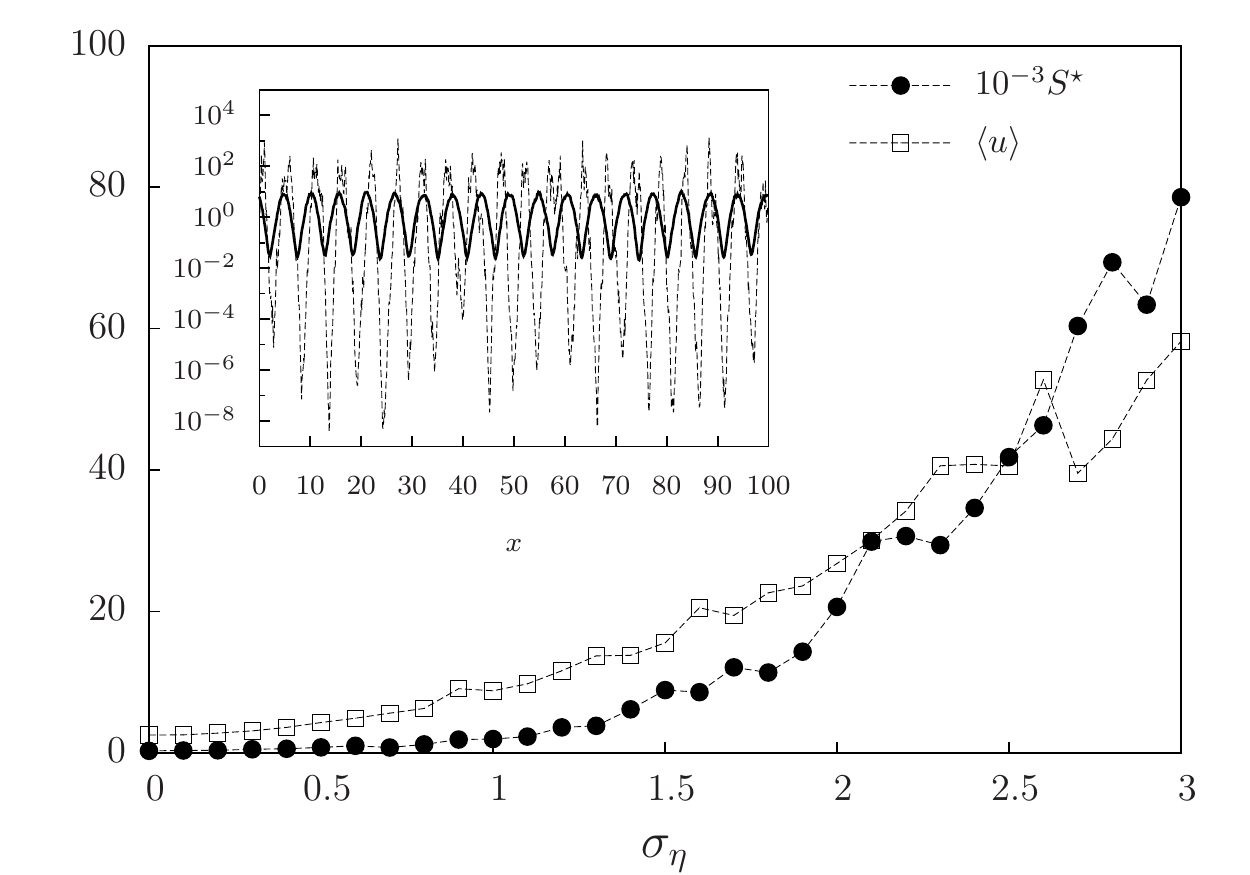}
\caption{Analysis of the spatial coherence for the choice:  
  $a=2$, $b=1$, $w=4$ $D=0.1$ and $\sigma_\xi=0$ (under the Stratonovich  prescription). 
The intensity of the dominant mode was scaled by a factor $10^{-3}$ just to use the same axis scale. 
In the inset we exhibit typical patterns for low (solid line) and high (dotted line) noise intensities, 
namely for $\sigma_\eta =$ 0.1 and 1.9, respectively. The dotted lines are  guides to the eyes.
}
\label{fig:noise_com_S}
\end{figure}

In the Stratonovich case, we observe a quite different behavior,  
as displayed in Fig.~\ref{fig:noise_com_S}. 
Increasing the noise intensity $\sigma_\eta$ induces growth both of the average 
level and of the intensity of the dominant mode, in such a way that also the ratio 
$S^*/\langle u \rangle$  increases.  
However, as shown in the inset of the same figure, while the amplitude of the patterns 
grows with increasing noise, 
their shape becomes more irregular, indicating that the other modes also grow together 
with the dominant one, as predicted by Eq.~(\ref{stocDispersion0}).

One can cast a Stratonovich stochastic differential equation into the form of an It\^o equation 
with an effective (or spurious) drift. 
For our Eq.~(\ref{nonlocalfisherlangevin}), this implies the change 
$ a \to a+\frac{K_\eta}{2} \sigma^2_\eta$. 
Because the additional term is positive, this change amounts to increasing the growth rate $a$. 
On the other hand, increasing $a$, with the other parameters fixed, does not alter 
the stability condition. 
This situation would lead to  increase   the average density and to strengthen patterns, 
which is in fact the outcome observed in Fig.~\ref{fig:noise_com_S},  
indicating that the destructive role observed for
It\^o noise (Fig.~\ref{fig:noise_com_I}) is not enough to spoil the constructive effect 
of the spurious drift.

\begin{figure}[h!]
\includegraphics[scale=0.6]{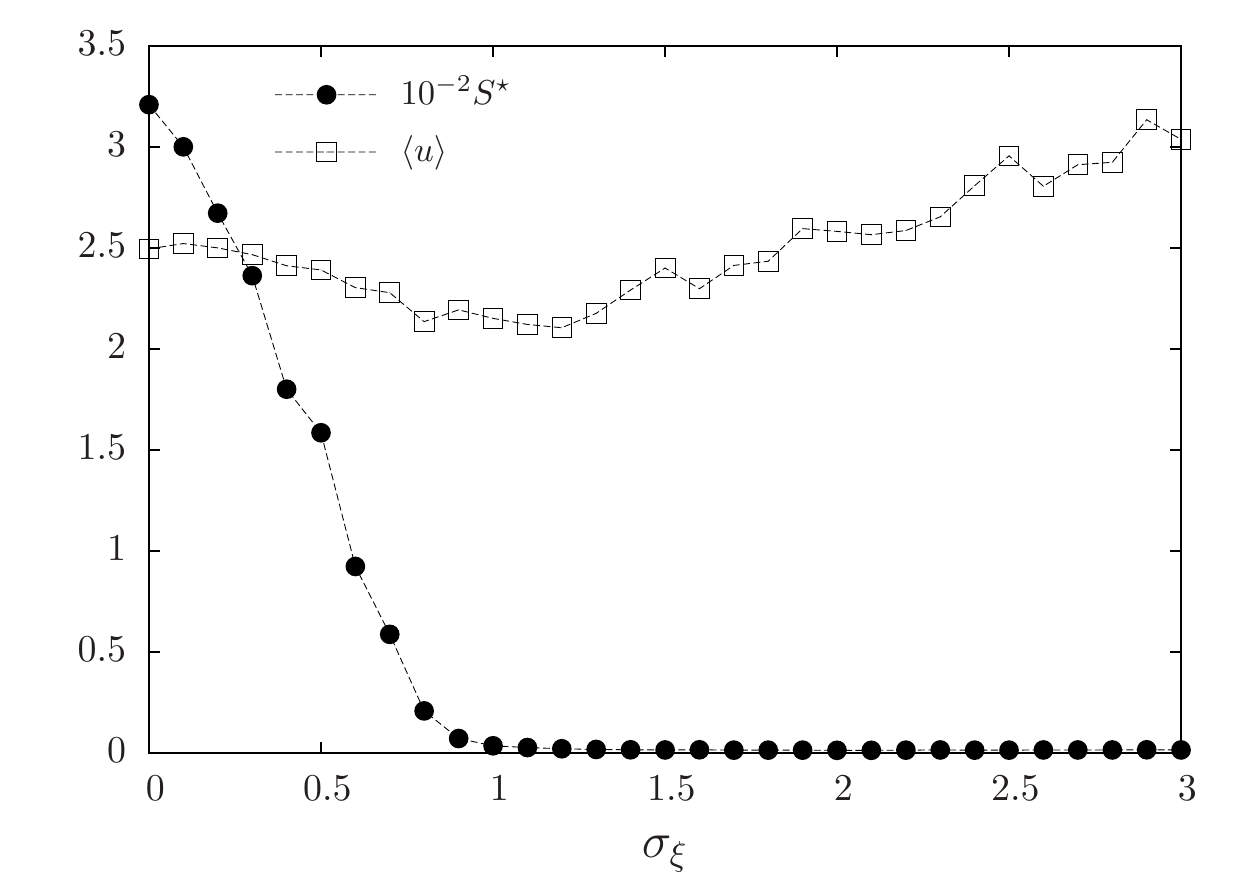}
\caption{Analysis of the spatial coherence for  
$a=2$, $b=1$, $w=4$, $D=0.1$ and $\sigma_\eta=0$.
The symbols correspond to numerical results.  The dotted lines are guides to the eyes.  
}
\label{fig:additive_com}
\end{figure}

The effect of additive white noise when the  multiplicative one is switched
off  ($\sigma_\eta= 0$) is shown in Fig.~\ref{fig:additive_com}. 
To ensure positiveness, negative fluctuations were trimmed,  by setting 
$u(x,t+dt)=0$, if  $u(x,t+dt)<0$. 
We did not apply any symmetrization procedure to keep the null mean value.
Then truncations rise the noise mean value, producing a consequent shift of the population average.
This  effect becomes visible for large noise intensities, where the probability of trimming 
is not negligible.

Like multiplicative It\^o fluctuations,  additive noise  also plays a destructive role in coherence,  
as can be observed in the decay of $S^\star$ with noise intensity.
However, in this case, there is not an extinction threshold and the average density remains finite.

\subsection{In the absence of deterministic patterns ($\lambda(k^\star)<0$)}
\label{sec:h}

In this subsection we concentrate in our main case of interest, 
that is when the homogeneous solution is stable despite nonlocality. 
The purpose of analyzing this situation is to verify 
if the introduction of noise in the dynamic rules can inject coherence.

\begin{figure}[h!]
\includegraphics[scale=0.6]{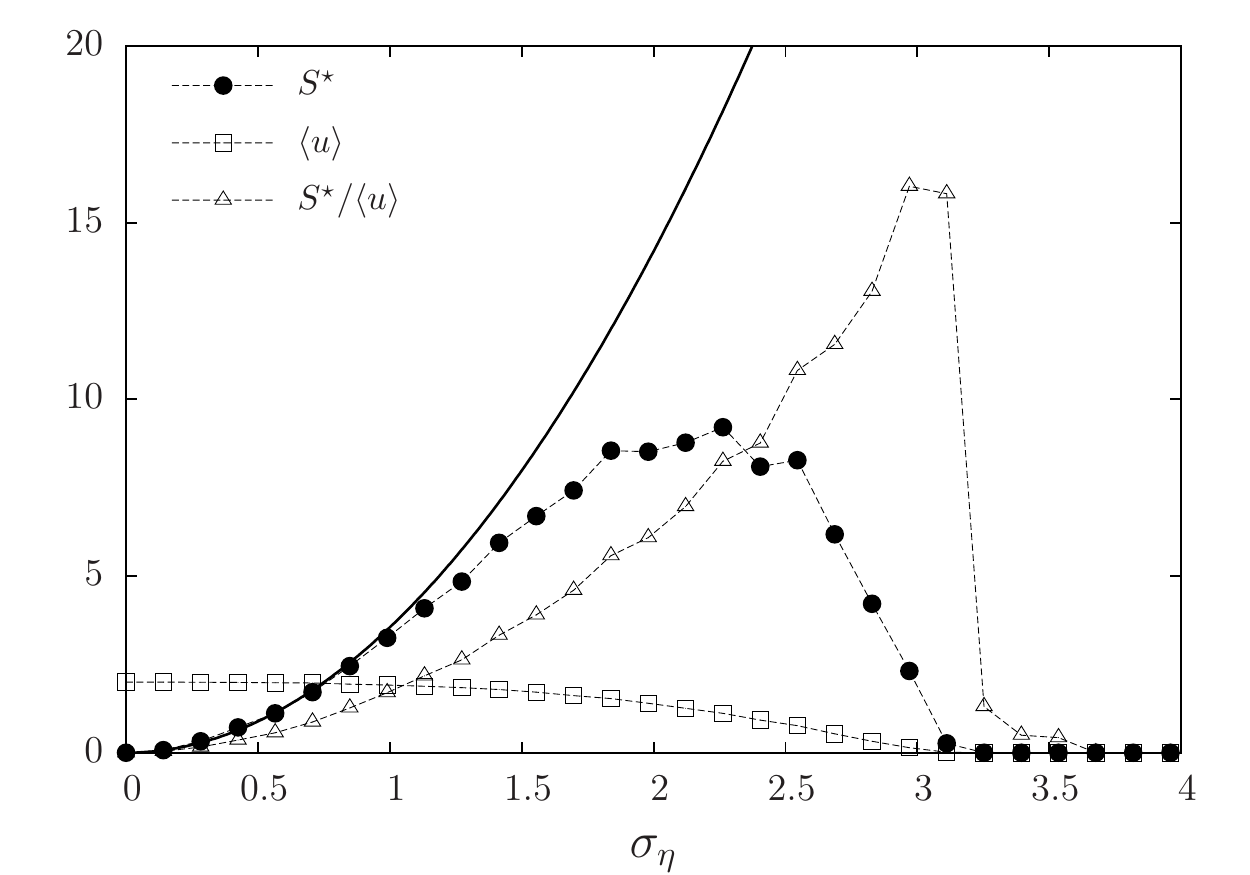}
\caption{Analysis of the spatial coherence for  
$a=2$, $b=1$, $w=4$, $D=1$ and $\sigma_\xi=0$, for It\^o noise. 
The symbols correspond to numerical results. The quotient $S^\star/\langle u \rangle$ is also plotted. 
The solid line  corresponds to the theoretical prediction  given by 
Eq.~(\ref{structure}), the dotted lines are guides to the eyes.  
}
\label{fig:plot_I}
\end{figure}

Let us first analyze the impact of the It\^o multiplicative noise.  
In Fig.~\ref{fig:plot_I}, we observe how the dominant mode intensity $S^\star$ changes 
as a function of the noise intensity $\sigma_\eta$.
Our results point out that when noise intensity is small enough ($\sigma_\eta < 1.0$),
 the increasing behavior $S^\star \propto \sigma_\eta^2$ predicted by Eq.~(\ref{structure}) occurs. 
However, when we increase the noise intensity beyond the linear regime, 
we note that there is a break in the monotonic behavior of $S^\star$ 
with a peak that characterizes an optimum value  $\sigma_\eta^{o}\approx 2.0$. 
Above this optimum value, noise starts to play a destructive role in spatial coherence. 
As a consequence, the dominant mode becomes less intense until it is completely destroyed. 
Actually, this is due to the concomitant decrease and extinction of the population, as shown 
by the quotient $S^\star/\langle u \rangle$ also exhibited in Fig.~\ref{fig:plot_I}.

\begin{figure}[h!]
\includegraphics[scale=0.8]{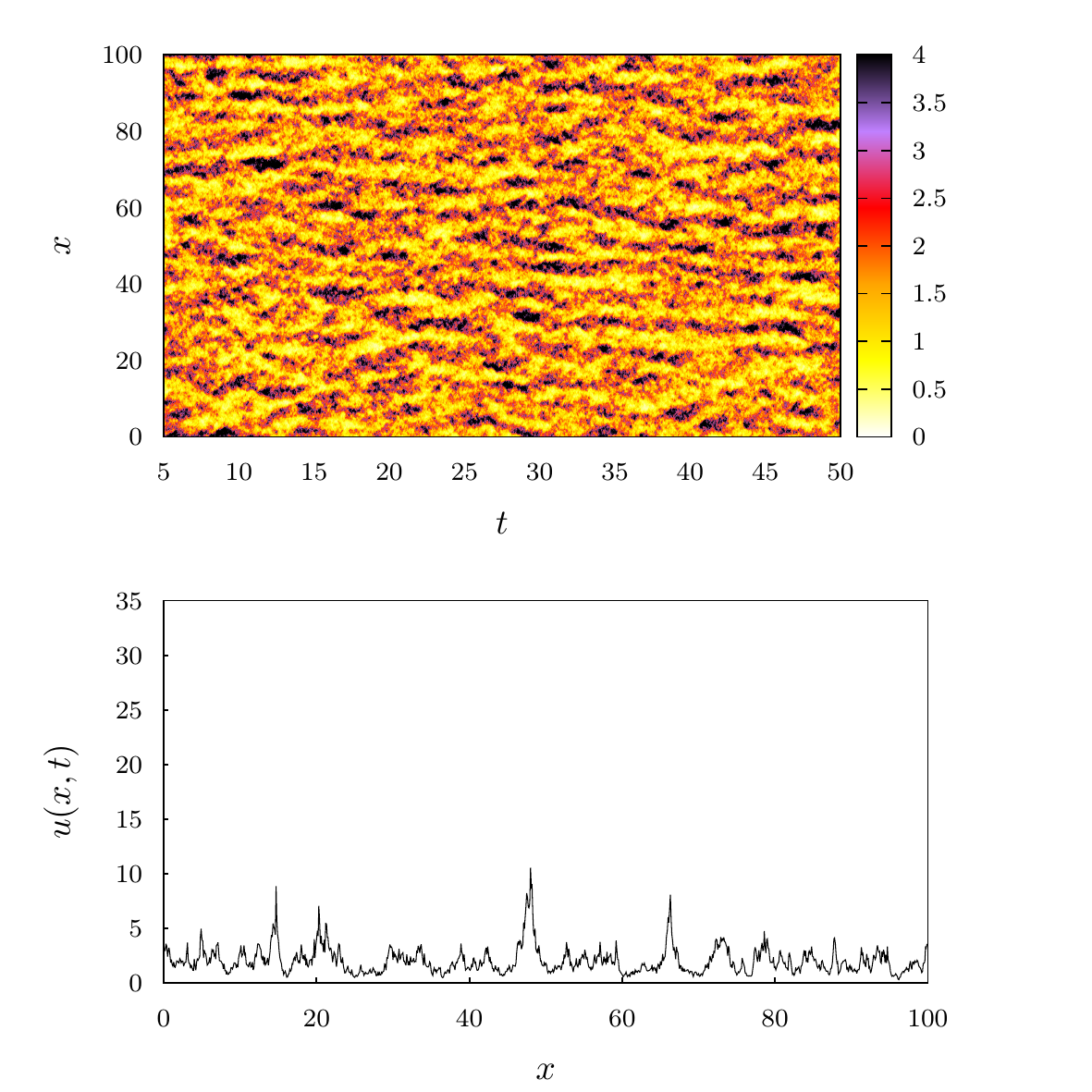}
\caption{(Color online) Time evolution of the density $u(x,t)$ in a color map (upper panel), 
for $a=2$, $b=1$, $w=4$, $D=1$, $\sigma_\xi=0$, $\sigma_\eta = 1.0$ (It\^o noise). 
In the lower panel we exhibit a density profile corresponding 
to a cut of the color map at $t=50$.  
}
\label{fig:evolution_I}
\end{figure}

On the one hand,  noise in the reproduction rate  affects the number of individuals in the population, as expected.
Also in this case, there exists a value $\sigma_{\eta}^c \approx 3$ that represents
 the \textit{extinction threshold}: if noise intensity exceeds that value, the population density 
vanishes, as discussed in Sec. \ref{sec:nonh}.  
On the other hand, although noise does not shift the dispersion relation, 
it forces an anticipation of mode instability, 
which is illustrated by the bursts of coherence displayed 
by density inhomogeneities in Fig.~\ref{fig:evolution_I}. 

\begin{figure}[h!]
\includegraphics[scale=0.6]{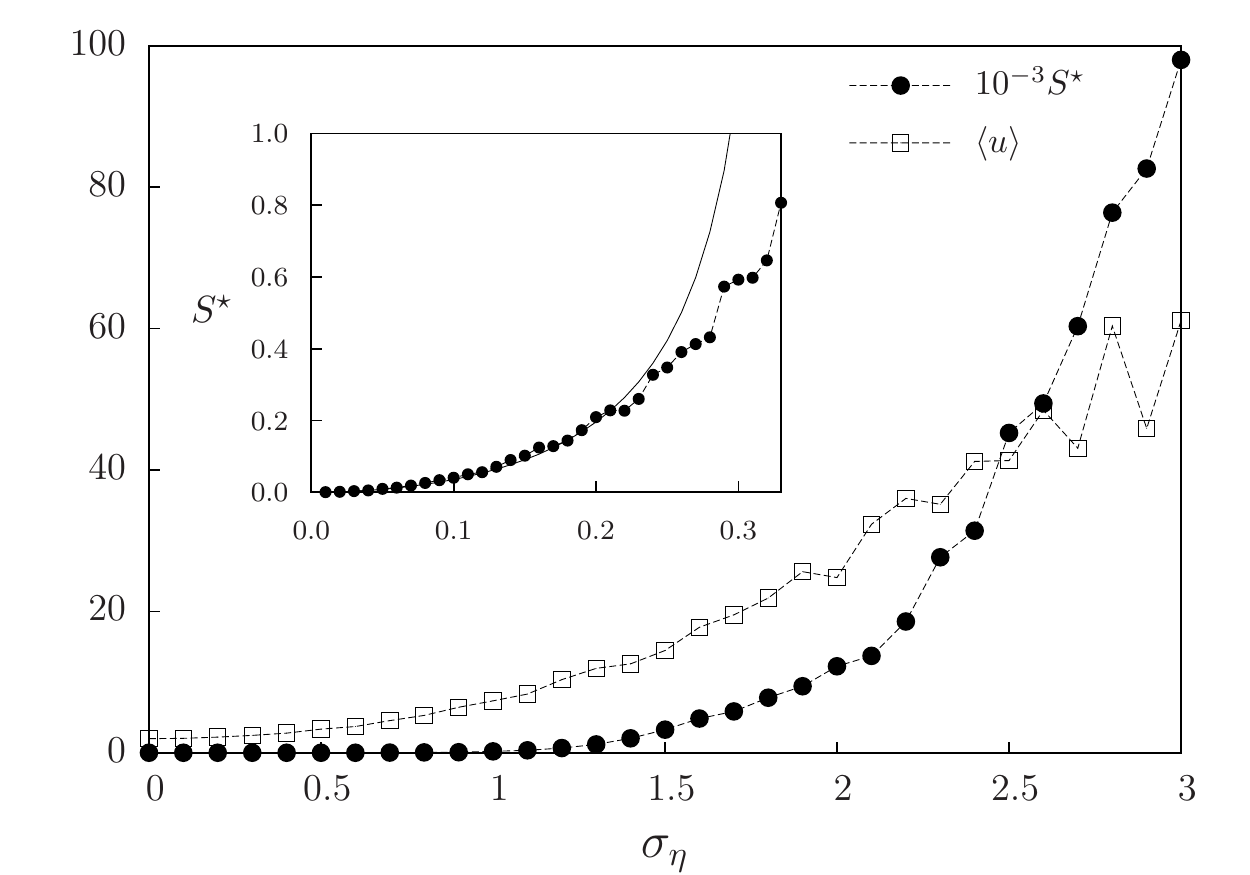}
\caption{Analysis of the spatial coherence for   $a=2$, $b=1$, $w=4$, $D=1$ 
and $\sigma_\xi=0$, for  Stratonovich noise. 
The symbols correspond to numerical results.  
The inset shows a blow up of the vicinity of the origin, where the solid line  
corresponds to the theoretical prediction given by Eq.~(\ref{structure}). 
}
\label{fig:plot_S}
\end{figure}

\begin{figure}[h!]
\includegraphics[scale=0.8]{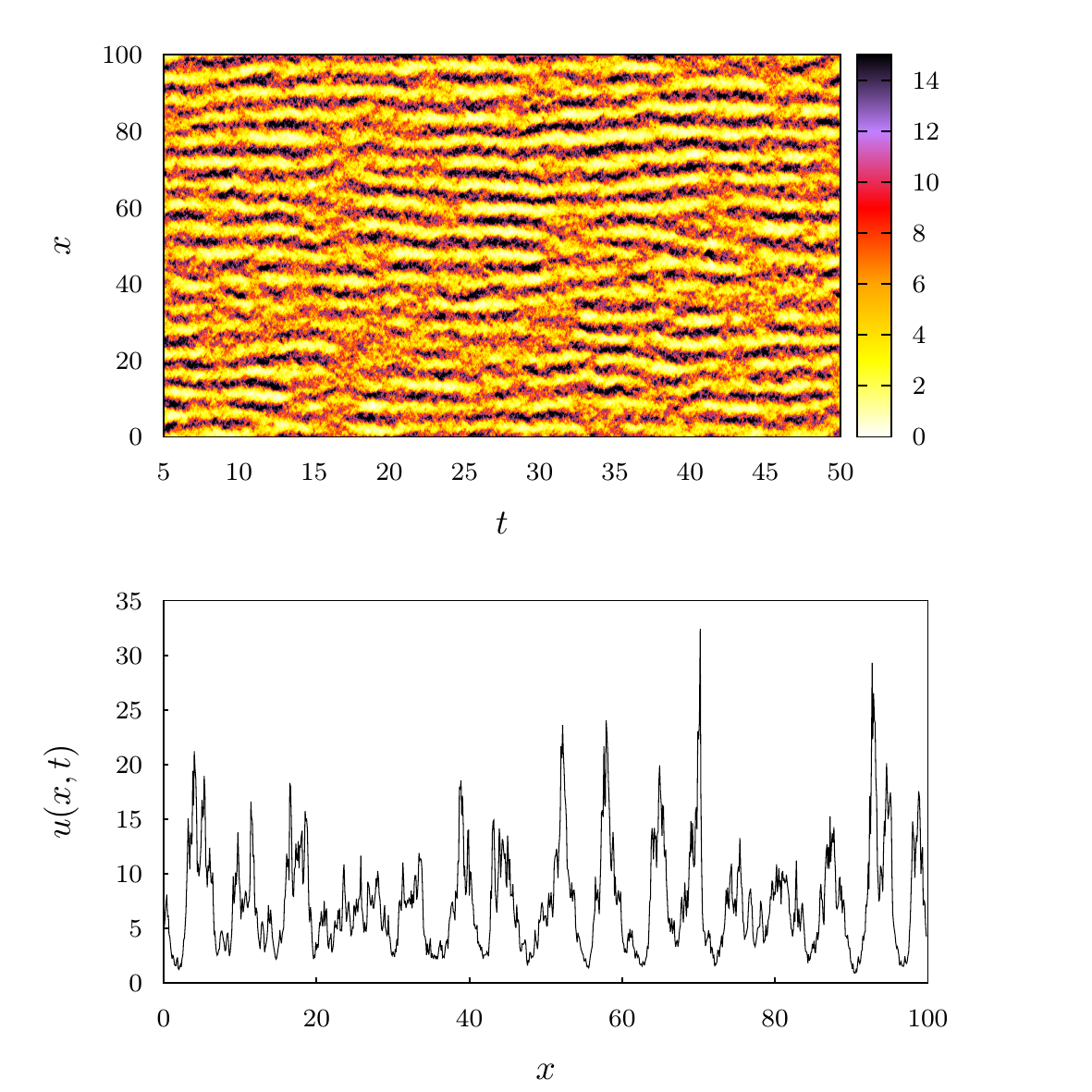}
\caption{(Color online) Time evolution of the density $u(x,t)$ in a color map (upper panel), 
for $a=2$, $b=1$, $w=4$, $D=1$, $\sigma_\xi=0$ and $\sigma_\eta = 1.0$ (Stratonovich case). 
In the lower panel we exhibit a profile corresponding to a cut of the color map at $t=50$.  
}
\label{fig:evolution_S}
\end{figure}

Now we perform the same analysis for the Stratonovich case.
Figure \ref{fig:plot_S} displays the analysis of spatial coherence, while the time evolution is 
depicted in Fig.~\ref{fig:evolution_S}. 
In the inset of the last figure we also show the theoretical prediction given by Eq.~(\ref{structure}), 
which is only valid up to a critical value, $\sigma_\eta^c \simeq 0.33$ in the case of the figure, 
point at which the theoretical structure function becomes divergent, although 
its numerical computation is  possible.

In terms of a spurious drift,  
Stratonovich noise would essentially lead to a larger growth rate, 
with the concomitant increase of the 
average density. Moreover, that spurious drift has the effect of shifting the 
dispersion relation, yielding Eq.~(\ref{stocDispersion0}). 
But, in contrast to  Sec. \ref{sec:nonh}, increasing noise intensity   
can shift  the maximum of the dispersion curve from the stability to the instability region 
for sufficiently large noise intensity (above its critical value).
When this happens,  
differently to the It\^o case for the same value of the parameters, 
persistence of spatial patterns emerges. 
The resulting profiles are similar to those observed for the parameter region in which 
patterns already  occur in the deterministic limit.

Although for both noises one has $S^\star>0$, 
indicating the presence of coherence, in the Stratonovich case, 
$\Lambda_1(k^\star) >0$ while in the It\^o case $\Lambda_0(k^\star) =\lambda(k^\star) <0$. 
That is, despite some kind of coherence is always revealed by noise, 
in the Stratonovich case there is 
persistence of the patterns, while in the It\^o case they are weakly correlated in time. 
Moreover, comparison of the profiles shown 
in Figs. (\ref{fig:evolution_I}) and (\ref{fig:evolution_S}), reveals a greater regularity 
and more pronounced peaks in the distribution $u(x,t)$ in the Stratonovich case.

\begin{figure}[!h]
\includegraphics[scale=0.6]{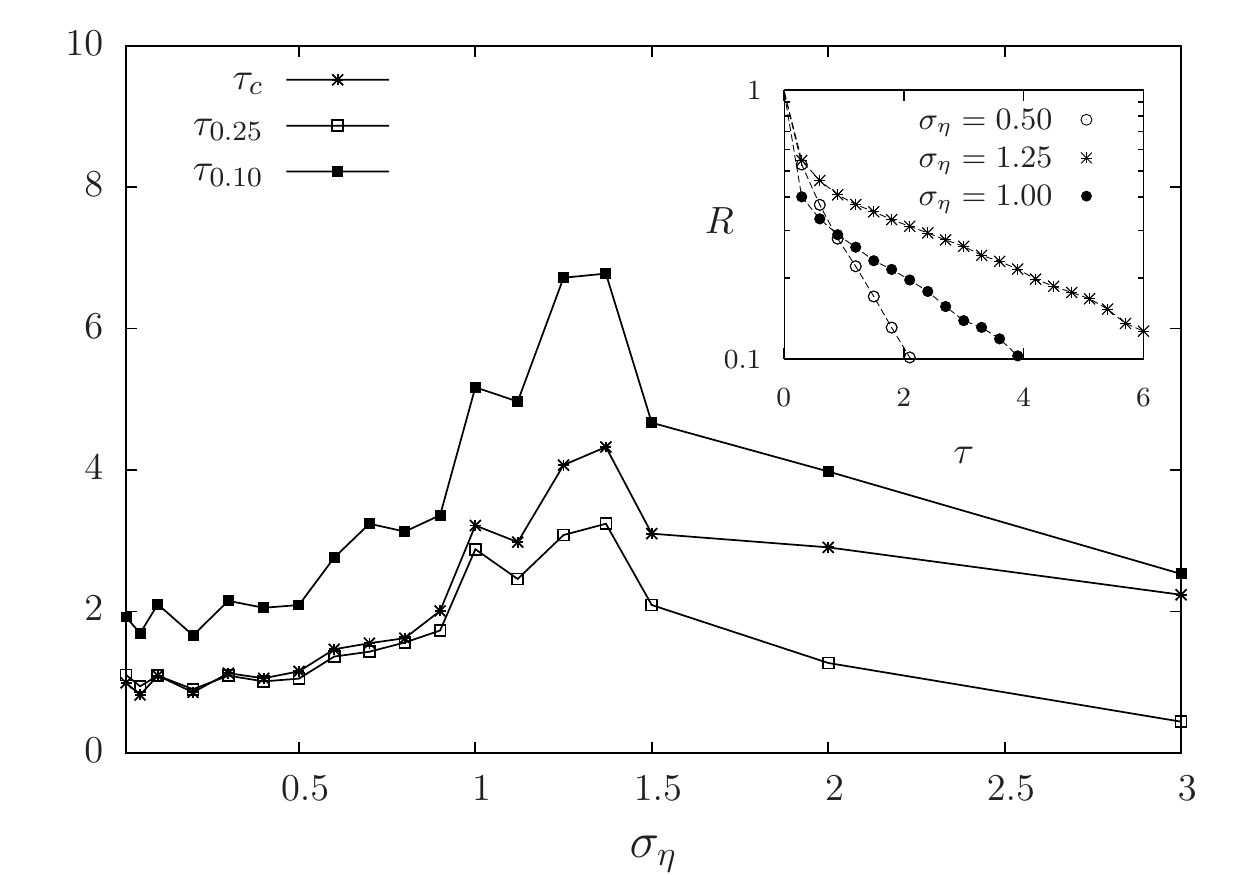}
\caption{Correlation times as a function of $\sigma_\eta$, 
for $a=2$, $b=1$, $w=4$, $D=1$, $\sigma_\xi=0$ (with Stratonovich noise).
The inset shows typical curves of the autocorrelation function vs the time lag $\tau$, 
from which correlation times were extracted: 
$\tau_c$ is the inverse rate of exponential decay, since there is an abrupt decay 
before the exponential regime, we also computed $\tau_{0.25}=\tau(R=0.25)$ and 
 $\tau_{0.10}=\tau(R=0.10)$.
}
\label{fig:persistence}
\end{figure}

In order to quantify the degree of persistence, we measured the spatial average of the 
time autocorrelation function of $u(x,t)$, $R(\tau)$ as a function of the time lag $\tau$. 
The autocorrelation function presents an exponential behavior after an abrupt decay, 
then, we considered different  effective  correlation times  
(as defined in Fig.~\ref{fig:persistence}), 
as measures of the degree of persistence. 
All these quantities plotted as a function of $\sigma_\eta$, 
under the Stratonovich interpretation, are presented in Fig.~\ref{fig:persistence}.
The figure shows that persistence first increases with noise intensity, attaining a maximum, 
and thereafter decays with larger noise intensities for which order is spoiled.  
Notice than in the limit of vanishing noise intensity, the correlation times do  not go to zero. 
The limiting values  remain  almost constant up to a value of $\sigma_\eta$ 
that approximately coincides with the critical one  
predicted by the condition $\Lambda_1(k^\star)=0$ in Eq.~(\ref{stocDispersion0}). 
In the case of the figure, the aforementioned critical value is $\sigma_\eta^c \simeq 0.33$. 
In fact, the kind of persistence observed in Fig.~\ref{fig:evolution_S} 
can be attributed to a positive maximum of the dispersion relation 
($\Lambda_\nu(k^\star)>0$). This condition is possible only for $\nu=1$ 
(Stratonovich interpretation) and $\sigma_\eta>\sigma_\eta^c$. 
Notice that,  under the It\^o interpretation ($\nu=0$), $\Lambda_0(k^\star)$ is always 
negative if $\lambda(k^\star)<0$, then such kind of persistent pattern cannot occur. 
In fact, for the It\^o simulations, we observed (not shown) that below the extinction threshold  
noise does not affect  the correlation times  that remain at the level of those 
at vanishing noise intensity in the  Stratonovich case. 
Hence we can conclude that for the same parameters, the effect of It\^o noise is equivalent 
to that of Stratonovich noise below $\sigma_\eta^c$.

\begin{figure}[h!]
\includegraphics[scale=0.6]{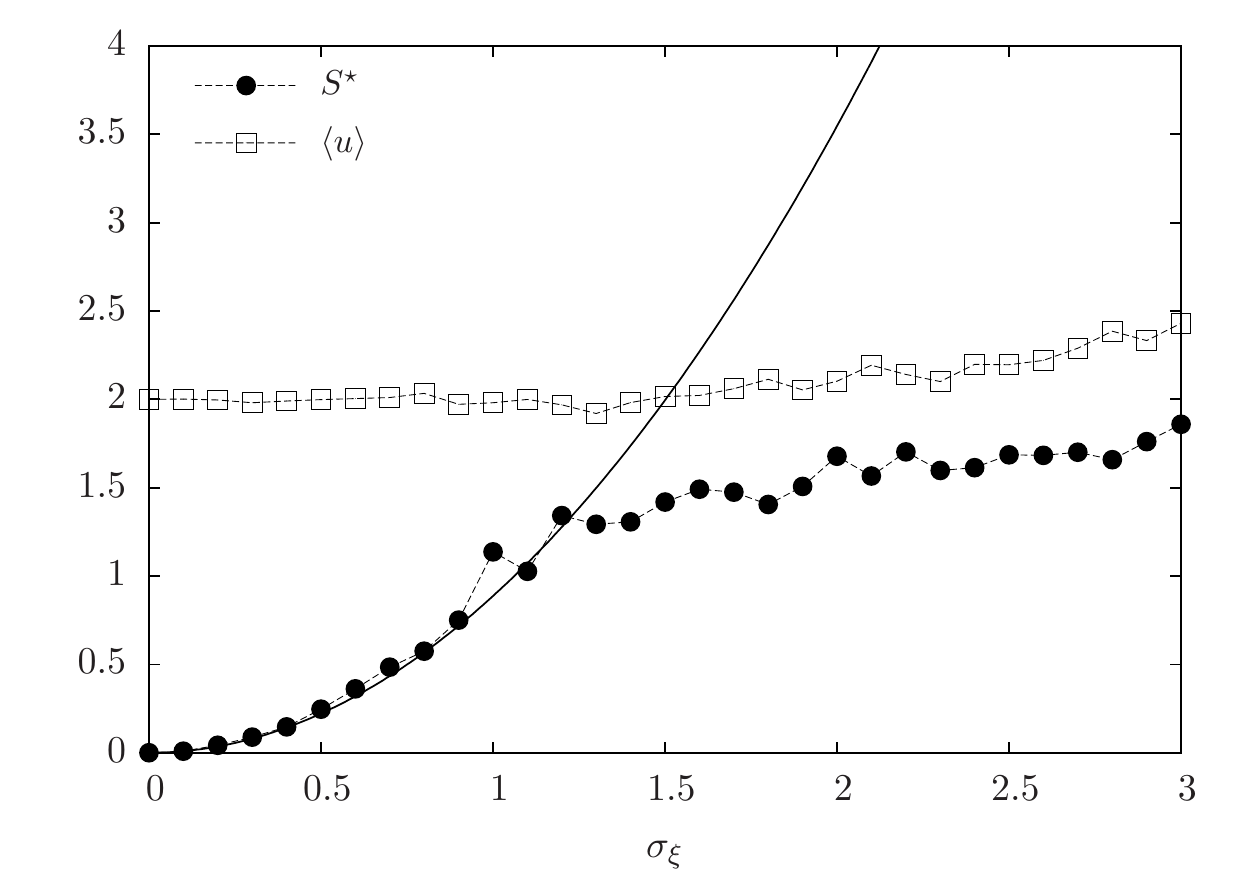}
\caption{Analysis of the spatial coherence for 
$a=2$, $b=1$, $w=4$, $D=1$ and $\sigma_\eta=0$.
The symbols correspond to numerical results. The solid line  corresponds to the theoretical prediction  given by 
Eq.~(\ref{structure}), the dotted lines are guides to the eyes.  
}
\label{fig:additive}
\end{figure}

The impact of additive noise is shown in Fig.~\ref{fig:additive}, confirming the possibility of 
induced coherence  in this case,  as predicted by Eq.~(\ref{structure}). 
However, patterns present  low temporal correlation,  similarly to   
those in  Fig.~\ref{fig:evolution_I} for It\^o fluctuations.
Notice that Eq.~(\ref{structure})   furnishes a good prediction of the impact of additive   fluctuations 
observed in numerical simulations.
This is because in the low noise intensity limit where Eq.~(\ref{structure}) holds, 
the necessity of applying the truncation rule seldom occurs.

\section{Final remarks}

We incorporated stochasticity in a generalized Fisher-KPP equation, 
namely by adding noise to a characteristic parameter, 
the growth rate, as well as by means of an additive noise term. 
Then, we focused on the impact of such fluctuations on the stability of the asymptotic state.  
Although there is a formal correspondence between the 
It\^o and Stratonovich rules to interpret multiplicative stochastic 
equations, we chose to present a parallel  between them, for the same set of parameters, 
as soon as they apply to different scenarios of environment fluctuations.

When patterns are deterministically induced, 
It\^o noise in parameter $a$ has a destructive role. 
Instead, when the homogeneous state is stable,  
noise destabilizes it,  
inducing the emergence of  potentially dominant modes, which are hidden in the noiseless limit. 
However,  extreme values of the noise intensity lead to extinction in both cases.

 On the other side, when noise is interpreted in the Stratonovich sense, 
multiplicative fluctuations are able to increase population size as well as coherence, 
be patterns deterministically induced or not. 
Moreover, a crucial difference, noticed from numerical simulations, 
is the persistence of spatial patterns in the Stratonovich case, 
which is absent in the It\^o one. 
We characterized the changes of persistence as a function of the intensity of Stratonovich noise, 
by directly measuring the temporal autocorrelation.

When patterns are deterministic, additive noise ($\sigma_\xi>0$) plays a destructive role in coherence. 
$S^\star$ decays with noise intensity, similarly to the scenario observed for 
 It\^o  multiplicative fluctuations, but there is not an extinction threshold. 
In the absence of deterministic patterns,  additive noise induces coherence, 
although with  low temporal correlation,  also as in the case of 
It\^o fluctuations.

Future perspectives would be to consider fluctuations in the other parameters 
and to assume that noises possess  characteristic finite scales, typical of  natural environments.

\vspace*{5mm}	
\section*{Acknowledgements}
C.A. and L.A.S. acknowledge partial financial support by Brazilian agency 
Conselho Nacional de Desenvolvimento Cient\'{\i}fico e Tecnol\'ogico (CNPq). 
L.A.S was also partially supported by Funda\c{c}\~ao de Amparo \`a 
Pesquisa do Estado de Sao Paulo (FAPESP).   
E.H.C. aknowledges financial support of Funda\c{c}\~ao de Amparo \`a 
Pesquisa do Estado do Rio de Janeiro (FAPERJ).

\end{document}